\documentclass[reprint,amsmath,amssymb,aps,prl,longbibliography]{revtex4-1}

\usepackage{graphicx}
\usepackage{dcolumn}
\usepackage{bm}
\usepackage{bbm}
\usepackage{todonotes}

\DeclareSymbolFont{Eulerscripteusm10}{U}{eus}{m}{n}
\SetSymbolFont{Eulerscripteusm10}{bold}{U}{eus}{b}{n}
\DeclareMathSymbol{\euW}{\mathord}{Eulerscripteusm10}{"57}
\DeclareMathSymbol{\euD}{\mathord}{Eulerscripteusm10}{"44}

\DeclareMathAlphabet{\mathbxit}{\encodingdefault}{\rmdefault}{bx}{it}   

\newcommand{\be}{\begin{eqnarray}}
\newcommand{\ee}{\end{eqnarray}}
\DeclareMathAlphabet{\pazocal}{OMS}{zplm}{m}{n}   

\newcommand{\ra}{\rangle}
\newcommand{\la}{\langle}
\newcommand{\tr}{\rm Tr}
\newcommand{\one}{\mathbbm 1}

\newcommand*{\at}{@}

\newcommand{\tht}{\tfrac{\theta_1}2}
\newcommand{\pht}{\tfrac{\theta_2}2}

\begin{document}


\title{Leggett-Garg inequalities cannot be violated in quantum measurements}

\author{Christoph Adami}
\email{adami\at msu.edu}
\affiliation{Beyond Center for Fundamental Concepts in Science\\
 Arizona State University, Tempe, AZ 85287}
\affiliation{Department of Physics and Astronomy, Michigan State University, East Lansing, MI 48824}


\begin{abstract}
Leggett and Garg derived inequalities that probe the boundaries of classical and quantum physics by putting limits on the properties that classical objects can have. Historically, it has been suggested that Leggett-Garg inequalities are easily violated by quantum systems undergoing sequences of strong measurements, casting doubt on whether quantum mechanics correctly describes macroscopic objects. Here I show that Leggett-Garg inequalities cannot be violated by any projective measurement. The perceived violation of the inequalities found previously can be traced back to an inappropriate assumption of non-invasive measurability. Surprisingly, weak projective measurements cannot violate the Leggett-Garg inequalities either because even though the quantum system itself is not fully projected via weak measurements, the measurement devices 
are. 
  
\end{abstract}

\maketitle


The Bell~\cite{Bell1964,Bell1966} and Leggett-Garg inequalities~\cite{LeggettGarg1985} probe quantum mechanical correlations in spatial and temporal dimensions, respectively. While Bell's inequality puts limits on the (classical) correlations between measurements of two systems that could be entangled with each other, the Leggett-Garg inequalities put limits on the correlation between measurement devices that measured the same quantum system consecutively. 
According to Leggett and Garg, these inequalities should be fulfilled by all systems that can be described by {\em macro-realistic theories}, namely those that fullfil two conditions: macrorealism and non-invasive measurability.
Many have argued that because the inequalities appear to be easily violated experimentally, the predictions of quantum mechanics violate realism, meaning that it is in doubt whether quantum mechanics can consistently and correctly describe macroscopic objects. Ballantine subsequently argued in a comment on Ref.~\cite{LeggettGarg1985} that instead it was the assumption of non-invasive measurability that contradicted quantum mechanics~\cite{Ballantine1987}, a criticism also leveled by Peres~\cite{Peres1989}. 

\begin{figure}[htbp] 
   \centering
   \includegraphics[width=0.5\textwidth]{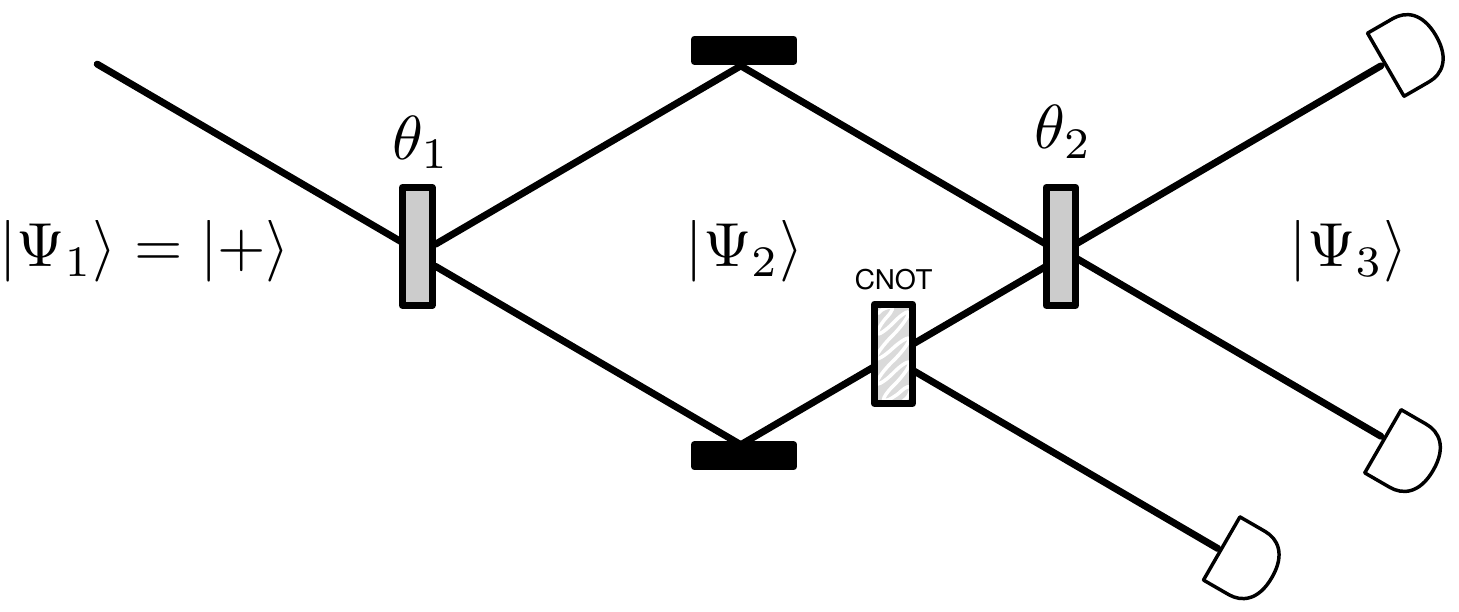} 
   \caption{Schematic view of the setup for two consecutive measurements on a prepared state $|+\rangle$.  
   \label{fig1}}
\end{figure}

Leggett and Garg's inequalities bound the expectation values of observables, measured consecutively on the same quantum system. It is easy to derive such inequalities for three consecutive measurements (the smallest number for which these inequalities can be derived) simply by insisting that the three measurement devices are consistent with each other. For example, for binary devices $A_1$, $A_2$ and $A_3$ that have outcomes $+$ and $-$ and a joint density matrix $\rho_{123}$ with diagonal elements $P(xyz)=\langle  xyz|\rho_{123}|xyz\rangle$ ($x$, $y$, and $z$ are the outcomes of measurement devices $A_1$, $A_2$ and $A_3$ respectively) and that is normalized according to \be
\sum_{xyz=-}^+P(xyz)=1\;,
\ee
the correlation function $K_{12}=\tr( \sigma_z\otimes\sigma_z\rho_{12})$ (for example) between the first two devices is the sum
\be
K_{12}&=&P(+\!+\!+)+P(+\!+\!-)-P(+\!-\!+)-P(+\!+\!-)\nonumber \\
           &-&P(-\!+\!+)-P(-\!+\!-)+P(-\!-\!+)-P(-\!+\!-)\;.\;\;\;\; \label{sum}
\ee 
In the definition of $K_{12}$, $\rho_{12}$ is the marginal density matrix of detectors $A_1$ and $A_2$, and $\sigma_z$ is the third Pauli matrix. Using the expression (\ref{sum}) (and the analogous ones for $K_{23}$ and $K_{13}$) it is easy to show that as long as the $P(xyz)$ are  probabilities, we can immediately derive three inequalities for the correlations
\be
B_1&=&K_{12}+K_{23}-K_{13}\leq1\;, \label{b1}\\
B_2&=&K_{12}+K_{13}-K_{23}\leq1\;,\label{b2}\\
B_3&=&K_{13}+K_{23}-K_{12}\leq1\;,\;\;\; \label{b3}
\ee
which are three of the four standard Leggett-Garg inequalities. Another common inequality, which in particular is used in~\cite{Kneeetal2012}, is
\be
f=K_{12}+K_{23}+K_{13}+1\geq0\;. \label{bf}
\ee
To study whether Leggett-Garg inequalities can be violated, I will first calculate the correlation functions $K_{ij}$ explicitly for the case where the initial state has been prepared as $|+\rangle$ (it is well-known that the correlators do not depend on the preparation of the initial state for the case of strong measurements). For this particular case, the correlation functions are
\be 
K_{12}&=&P(+\!+\!+)+P(+\!+\!-)-P(+\!-\!+)-P(+\!-\!-)\;,\ \ \ \ \  \label{k12}\\
K_{13}&=&P(+\!+\!+)-P(+\!+\!-)+P(+\!-\!+)-P(+\!-\!-)\;,\ \ \ \ \ \label{k13}\\
K_{23}&=&P(+\!+\!+)-P(+\!+\!-)-P(+\!-\!+)+P(+\!-\!-)\;. \ \ \ \ \ \label{k23}
\ee
To calculate the four functions $P(+yz)$, we only have to perform two consecutive measurements on the preparation $|\Psi_1\rangle=|+\rangle$ at angles $\theta_1$ and $\theta_2$ with respect to the preparation~\footnote{This is because the preparation can be viewed as the result of the first measurement.}. We can schematically set up the experiment as in Fig.~\ref{fig1}, where the states $|\pm\rangle$ can be understood as path variables in a Mach-Zehnder interferometer (for example,~\cite{Emaryetal2012}) and the ``beam splitter" implements a unitary rotation~\footnote{I ignore a possible phase $\phi$ in this rotation since the perceived violation of the Leggett-Garg  inequalities is maximal at $\phi=0$ anyway.}
\be
|+\ra&=&\cos(\tht)|\theta_1\ra-\sin(\tht|\bar\theta_1\ra\\
|-\ra&=&\sin(\tht)|\theta_1\ra+\cos(\tht)|\bar\theta_1\ra\;,
\ee
which defines $|\theta_1\rangle$ and $|\bar\theta_1\rangle$ implicitly. 
The first measurement is implemented using the unitary operator that measures the quantum state in the rotated $(\theta_1,\bar\theta_1)$ basis and conditionally flips the ancilla associated with this measurement: a controlled-NOT (CNOT) operation
\be
U_1=|\theta_1\rangle\langle \theta_1|\otimes \one +|\bar\theta_1\ra\la \bar\theta_1|\otimes \sigma_x\;, \label{CNOT}
\ee 
where $\sigma_x$ (the first Pauli matrix) flips the ancilla. This measurement is indicated by the CNOT operator in Fig.~\ref{fig1}.
Before the measurement, the preparation is described by the wave function $|\Psi_1\ra=|+\ra|0\rangle_1|0\rangle_2$, where $|0\ra_1$ is an ancillary state that represents the memory of the state preparation (this will turn out to be convenient for bookkeeping purposes later on), and  $|0\ra_2$ is the ancilla that will record the first measurement. 

After the measurement implemented by $U_1$, we are left with the wave function $|\Psi_2\ra=U_1|\Psi_1\ra$:
\be
|\Psi_2\ra=\cos(\tht)|\theta_1\rangle|0\rangle_1|0\rangle_2-\sin(\tht)|\bar\theta_1\rangle|0\rangle_1|1\rangle_2\;.\ \ \ \ \ 
\label{psi2}
\ee
The operation (\ref{CNOT}) implements a strong projective measurement on the quantum state and ancilla. Alternatively we can perform a {\em weak} measurement~\cite{Aharonovetal1988,Ritchieetal1991,Lundeenetal2011}, using instead
$U_1=e^{-iH(g)}$ with the interaction Hamiltonian $H(g)=gP_\theta\otimes \sigma_y$ with $P_\theta=|\theta_1\ra\la\theta_1|$ and where $\cos(g)=\sqrt{1-\epsilon^2}$. Weak measurements do not flip the ancilla to its orthogonal state, but rather move it by a smaller angle to the state 
$|\epsilon\rangle_2=\sqrt{1-\epsilon^2}|0\rangle_2+\epsilon|1\rangle_2$. In the worst case ($\epsilon=0$) the measurement device is insensitive to the quantum state, that is, no measurement is performed.

The wave function $|\Psi_2\rangle$ after a weak measurement is simply obtained from (\ref{psi2}) by replacing $|1\rangle_2\to |\epsilon\rangle_2$, that is, replacing the ancillary (which in a strong measurement is in a known state) by a superposition, giving rise to fuzziness in the measurement outcome~\cite{Aharonovetal1988,Lundeenetal2011,LundeenBamber2012,Dresseletal2014,Glick2017,Curic2018}. Clearly, the strong measurement returns in the limit $g\to\pi/2$ $(\epsilon\to1)$. 

The wave function $|\Psi_2\rangle$ is sufficient to calculate the density matrix of the pair of measurement devices by tracing out the quantum state
\be
\rho_{12}=|0\ra_1\la0|(\cos^2(\tht)|0\ra_2\la0|+\sin^2(\tht)|\epsilon\ra_2\la\epsilon|) \ \ \ \ \ \ \ \ 
\ee 
and hence (using $\la \epsilon|\sigma_z|\epsilon\ra=1-2\epsilon^2$)
\be 
K_{12}=\tr(\rho_{12}\sigma_z\otimes\sigma_z)=(1-\epsilon^2) +\epsilon^2\cos(\theta_1)\;. \label{12}
\ee

In the limit of strong measurements ($\epsilon=1$) expression~(\ref{12}) recovers the standard result $K_{12}=\cos(\theta_1)$, found previously by many authors~\cite{KoflerBrukner2008,Gogginetal2011,Athalyeetal2011,Kneeetal2012,Emaryetal2012}, see ~\cite{Emaryetal2014} for a more complete list. 

We are now ready to perform the second (strong) measurement, with the device set an angle $\theta_2$ with respect to the first one.   To do this, we rewrite the quantum system's basis states, (currently written in the $\theta_1$-basis) in the $\theta_2$ basis instead :
\be
|\theta_1\ra&=&\cos(\pht)|\theta_2\ra-\sin(\pht)|\bar\theta_2\ra\;,\\
|\bar\theta_1\ra&=&\sin(\pht)|\theta_2\ra+\cos(\pht)|\bar\theta_2\ra\;.
\ee
With a third qubit ancilla's basis states $|0\ra_3$ and $|1\ra_3$, the wave function after the third measurement using $U_3=|\theta_2\ra\la\theta_2|\otimes\one+|\bar\theta_2\ra\la\bar\theta_2|\otimes \sigma_x$ (with $\sigma_x$ acting on the third qubit's Hilbert space) becomes 
\begin{widetext}
\be
|\Psi_3\ra&=&|\theta_2\ra\biggl(\!\Bigl[\cos(\tht)\cos(\pht)|0\ra_2 -\sin(\tht)\sin(\pht)|\epsilon\rangle_2\Bigr)\rangle|0\rangle_3\nonumber\\ 
&-&|\bar\theta_2\rangle\Bigl[\cos(\tht)\sin(\pht)\ra|0\ra_2+\sin(\tht)\cos(\pht)|\epsilon\ra_2\Bigr)|1\ra_3
\label{psi3}
\ee 
\end{widetext}
From this expression we can obtain \begin{widetext}
\be
\rho_{23}&=&\tr_2(|\Psi_3\rangle\la\Psi_3|)
=|0\ra_2\la 0|\Bigl(\cos^2(\pht)|0\ra_3\la0|+\sin^2(\pht)|1\ra_3\la1|\Bigr) 
+|\epsilon\ra_2\la\epsilon|\Bigl(\sin^2(\pht)|0\ra_2\la 0|+\cos^2(\pht)|1\ra_3\la 1|\Bigr)
\ee
\end{widetext}
which immediately yields 
\be
K_{23}=(1-\epsilon^2)\cos(\theta_1)\cos(\theta_2)+ \epsilon^2\cos(\theta_2)\;.\label{23}
\ee
For strong measurements ($\epsilon=1)$ this result mirrors $K_{12}$ as a function of the angle $\theta$, and it is obvious why they must be the same in that case: for pairs of consecutive strong measurements, the result does not depend on what happened before. It's also clear that for weak measurements this is no longer true, and history (the influence of $\theta_1$) matters. 

Let's now calculate $K_{13}$ directly from the wave function (\ref{psi3}). Again for an initial state $|+\rangle$, we obtain
\be
K_{13}=\cos(\theta_1)\cos(\theta_2) \label{k13a}
\ee
for any $\epsilon$ (but the result will depend on $\epsilon$ for arbitrary preparations as we will see later on). We note immediately that $K_{13}$ is the same as the value of $K_{23}$ in the limit of no first measurement ($\epsilon=0$) as it should, because after all if the first measurement did not take place, then for this state preparation the second device (which records the first measurement, as the first device records the preparation) simply is a mirror of the first device. 

But (\ref{k13a}) differs from the result usually cited in the literature, namely $K_{13}=\cos(\theta_1+\theta_2)$. The latter result would follow if the first measurement was zero-strength ($\epsilon=0$), and most authors simply argue that a non-invasive measurement is tantamount to not carrying it out. However, this is clearly wrong.  Eq.~(\ref{psi3}) shows that quantum measurements  are {\em always} invasive unless $\epsilon=0$, that is, the measurement is not carried out. This is precisely what Ballantine~\cite{Ballantine1987} and Peres~\cite{Peres1989} pointed out more than thirty years ago (along with, even earlier, Dicke~\cite{Dicke1981})

I will discuss this in more detail momentarily, but let us first study the Leggett-Garg inequalities (\ref{b1}-\ref{b3}) using these results. For the special case $\theta_1=\theta_2\equiv \theta$ I find
\be
B_1 & =& 1-\epsilon^2(1-\cos(\theta))^2\;,\\
B_2 &=& 1-\epsilon^2\sin^2(\theta)\;,\\
B_3 & = &-1+\epsilon^2\sin^2(\theta)\;,
\ee
which are all bounded from above by 1, that is, the Leggett-Garg inequalities cannot be violated by either strong or weak measurements. Let's now test how these inequalities depend on the state preparation.

Prepare an initial quantum state as a superposition $|\Psi_0\ra=\alpha|+\ra+\beta|-\ra$ (with $|\alpha|^2+|\beta|^2=1$). I then find more generally
\be
K_{12}&=&(|\alpha|^2-|\beta|^2)(1-\epsilon^2)+\epsilon^2\cos(\theta_1)\;, \label{ka}\\
K_{23}&=&(|\alpha|^2-|\beta|^2)(1-\epsilon^2)\cos(\theta_1)\cos(\theta_2)+\epsilon^2\cos(\theta_2)\;,\ \ \ \ \\
K_{13}&=&\cos(\theta_1)\cos(\theta_2)-2\sqrt{1-\epsilon^2}|\alpha||\beta|\sin(\theta_1)\sin(\theta_2)\;,\ \ \ \ \ 
\label{kc}
\ee
which recovers the previous results for $|\alpha|^2=1$. 

The correlation functions (\ref{ka}-\ref{kc}) are interesting from the point of view of weak measurements, because they imply that the correlation between subsequent measurements (in which one of them is weak) has a component that is due entirely to the state preparation (the term proportional to $1-\epsilon^2$), and a term due to the angle between measurements. If the weak measurement is first, the state preparation term is also modulated by the relative angles. 

In the special case $|\alpha|=1/\sqrt2$ the state preparation term in the correlation function disappears, and we  obtain the simple equations
\be
K_{12}&=&\epsilon^2\cos(\theta_1)\;, \label{k12b}\\
K_{23}&=&\epsilon^2\cos(\theta_2)\;,\\ \label{k23b}
K_{13}&=& \cos(\theta_1)\cos(\theta_2)-\sqrt{1-\epsilon^2}\sin(\theta_1)\sin(\theta_2)\;. \label{k13b}
\ee
For truly non-invasive first measurements (that is, first measurements that are not performed) $K_{13}=\cos(\theta_1+\theta_2)$, the result usually cited in the literature for non-invasive three-point measurements (counting the state preparation as the first measurement). But note that in the limit $\epsilon\to0$ (choosing two-point measurements) the correlation functions $K_{12}$ and $K_{23}$ {\em disappear} as they must (there is no contribution from the state preparation) so that again the Leggett-Garg inequalities cannot be violated, even in the limit $\epsilon=0$. Indeed, it now becomes clear that the alleged violation of Leggett-Garg inequalities (both experimental and theoretical) is a consequence of 
using expressions mixing strong and weak measurements: forgoing the first measurement (by assuming it is non-invasive and thus not performing it) is equivalent to choosing $\epsilon=0$. If this is your choice, you cannot assume that $\epsilon=1$ when writing down (or measuring) $K_{12}$ and $K_{23}$ in separate experiments where you do carry out the first measurement.

Using the general expressions (\ref{ka}-\ref{kc}), the inequalities become
\begin{widetext}
\be
B_1&=&1-\epsilon^2(1-\cos(\theta_1))(1-\cos(\theta_2))-2|\beta|^2(1-\epsilon^2)(1-\cos(\theta_1)\cos(\theta_2))+2|\alpha||\beta|\sqrt{1-\epsilon^2}\sin(\theta_1)\sin(\theta_2)\leq1\;,\ \ \ \\
B_2&=&1-\epsilon^2(1+\cos(\theta_1))(1-\cos(\theta_2))-2|\beta|^2(1-\epsilon^2)(1-\cos(\theta_1)\cos(\theta_2))-2|\alpha||\beta|\sqrt{1-\epsilon^2}\sin(\theta_1)\sin(\theta_2)\leq1\;,\ \ \ \\
B_3&=&1-\epsilon^2(1-\cos(\theta_1))(1+\cos(\theta_2))-2|\alpha|^2(1-\epsilon^2)(1-\cos(\theta_1)\cos(\theta_2))-2|\alpha||\beta|\sqrt{1-\epsilon^2}\sin(\theta_1)\sin(\theta_2)\leq1\;,\ \ \ 
\ee
\end{widetext}
It is straightforward to show that these inequalities cannot be violated for any state preparation or any strength of projective measurement, which implies that quantum mechanics describes macroscopic objects perfectly adequately. 

In the last ten years several articles have appeared that wondered why it was so easy to violate the Leggett-Garg inequalities for strong measurements (see, e.g.,~\cite{KoflerBrukner2008}) even though the measurement devices are clearly classical. Kofler and Brukner~\cite{KoflerBrukner2008} offered two alternatives to escape this dilemma: Either the ``quantumness" of the measurement devices quickly decoheres, or else it is in fact impossible to perform ``sharp" (that is, strong) measurements. We see now that the resolution is an entirely different one: the inequalities are never violated, no matter how fuzzy the measurements are. 

That the assumption of non-invasive measurability is at the origin of the perceived violation of the Leggett-Garg inequalities has actually been demonstrated experimentally (albeit it inadvertently)~\cite{Katiyaretal2013}. Katiyar et al.\ measured both the three-point function $P(x,y,z)$ as well as the two-point function $P(x,z)$, and noted that ``the grand probability [$P(x,y,z)$] cannot reproduce the two-time joint probabilities as the marginals"~\cite[p. 052102-5]{Katiyaretal2013}. The authors concluded that ``the grand probability is not legitimate in the quantum case" when on the contrary it is the assumption of non-invasiveness that is not legitimate. Using the actual measured $P(x,y,z)$ in that experiment surely would demonstrate that the Leggett-Garg inequalities are inviolate. 

An experiment that claims to have demonstrated the most convincing violation of the inequalities, using ``ideal non-invasive" measurements on spin-bearing phosphorus impurities in silicon, suffers from the same exact problem. Knee et al.~\cite{Kneeetal2012} assumed that not performing the intermediate measurement when obtaining $K_{13}$ was warranted because the measurements performed by the team on the first (of two) ancilla for $K_{12}$ and $K_{23}$ (in a separate experiment) was non-invasive.  
Indeed, the experiments in~\cite{Kneeetal2012} were cleverly designed in such a way that half of the measurements (at the ``middle" position for $K_{12}$ and $K_{23}$) were performed using a unitary operator as in (\ref{CNOT}), and half with the ``anti-CNOT" operation
\be
\bar U_2=
|\theta_1\ra\la\theta_1|\otimes \sigma_x +  |\bar\theta_1\ra\la\bar\theta_1|\otimes \one\label{baru2}\;.
\ee
In those experiments, when determining $K_{12}$ and $K_{23}$ the cases where the ``flip" was detected would be discarded, so that the ancilla was left undisturbed in the kept cases (hence the claim of non-invasiveness). Those are strong measurements, and therefore $\epsilon=1$ must be used there. For $K_{13}$ instead, the authors assumed $\epsilon=0$ for the middle measurement. The present analysis predicts that if the authors would re-analyze their data using the ``opposite" data sets (using the discarded set instead of the kept one) the result would be entirely unchanged, that is, they would find a violation of the Leggett-Garg inequalities even though data from a fully invasive experiment was used. Manipulating perceived invasiveness by conducting ``interaction-free" measurements has no bearing on the correlation functions.
This teaches us (once more) that the state of the ancillary is in general not diagnostic of the state of the quantum system, which will be disturbed even when the measurement device is not (unless $\epsilon=0$)~\cite{Dicke1981,CerfAdami1998,AdamiCerf1999,GlickAdami2017,GlickAdami2017c}.  If the first measurement were to be performed instead, there is no doubt that the authors would find $K_{13}=\cos(\theta_1)\cos(\theta_2)$, and the Leggett-Garg inequalities inviolate.

The conclusion that Leggett-Garg inequalities cannot be violated by strong or weak projective measurements seems to contradict experimental evidence of just such a violation for weak measurements by Goggin et al.~\cite{Gogginetal2011}. However, in the setup used by those authors the first and second measurements are performed in parallel (and on different observables) rather than consecutively, and formally represent a variant of the CHSH inequalities~\cite{Clauseretal1969} instead, rather than Leggett-Garg inequalities (as pointed out also in \cite{Emaryetal2014}).  The perceived violation of an entropic version of the Leggett-Garg inequalities in theory~\cite{UshaDevietal2013} and in experiment~\cite{Katiyaretal2013} also suffers from mixing statistics from weak and strong measurements, as I point out elsewhere~\cite{Adami2019a}. 

While in hindsight the Leggett-Garg inequalities should never have been proposed as a test of macrorealism (a theory that implies that quantum mechanics cannot consistently describe macroscopic objects) because the inequalities concern only the classical measurement devices and not the quantum system itself, they are obviously useful in probing the relative state of measurement devices that consecutively measured the same quantum state. What these inequalities teach us then is not that quantum mechanics does not adequately describe classical objects. They instead teach us that classical objects cannot adequately describe quantum objects.





\begin{acknowledgments}
I would like to acknowledge the hospitality of the Beyond Center for Fundamental Concepts in Science at Arizona State University, where this research was accomplished, as well as thank Paul Davies and G. Andrew Briggs for discussions. The work was funded in part by the John Templeton Foundation. 
\end{acknowledgments}

\bibliography{quant-PRL}

\end{document}